%% file: main.tex
\documentclass[a4paper,11pt]{article}
\usepackage{pos}
\usepackage{caption}
\usepackage{subcaption}

\usepackage{amsmath,amssymb}
\usepackage{xspace}

\usepackage{wrapfig}
\usepackage{graphicx}
\usepackage{enumitem}

\let\oldbibitem\bibitem
\renewcommand{\bibitem}[1]{\oldbibitem{#1}\vspace{-5pt}}  % adjust as needed

\title{First Data of the 3000\,km$^2$ Radio Detector at the Pierre Auger Observatory}
\ShortTitle{First Data of the 3000\,km$^2$ Radio Detector at the Pierre Auger Observatory}

\author*[a]{Bjarni Pont}

\affiliation[a]{Department of Astrophysics/IMAPP, Radboud University, P.O.\ Box 9010, NL-6500 GL Nijmegen, The Netherlands}

\onbehalf{for the Pierre Auger Collaboration$^b$}
\affiliation[b]{Observatorio Pierre Auger, Av.\ San Mart{\'\i}n Norte 304, 5613 Malarg\"ue, Argentina\\
Full author list: {\rm\url{https://www.auger.org/archive/authors_icrc_2025.html}}}

\emailAdd{spokespersons@auger.org}

\abstract{In this contribution, we present the status and first data from the Radio Detector (RD) at the Pierre Auger Observatory, consisting of $1660$ radio antennas deployed across the $3000$\,km$^2$ surface detector array. These antennas measure the radio emission from extensive air showers in the $30-80$ MHz band, enabling electromagnetic energy measurements for air showers with zenith angles above $65$\,$^\circ$. Combined with the muonic measurements from the water-Cherenkov detectors (WCDs), this allows mass composition studies at the highest energies. The large-scale deployment of the RD began in November 2023 and was completed in November 2024. A full end-to-end calibration shows consistency between Galactic and in-lab calibration to better than $5$\% and includes continuous monitoring for hardware failures, ensuring, for example, antenna alignment within $5$\,$^\circ$. We present the first data, demonstrating a strong correlation between the electromagnetic energy measured by the RD and the total shower energy measured by the WCD, confirming that the detector chain—including triggering, data readout, absolute calibration, and reconstruction is well understood. We highlight a particularly impressive $32$\,EeV shower at a zenith angle of $85$\,$^\circ$, producing a $50$\,km-long radio footprint, showcasing the unique capabilities of this detector.}

\ConferenceLogo{Pos_ICRC2025_logo}

\FullConference{%
39th International Cosmic Ray Conference (ICRC2025)\\
 15–24 July 2025\\
Geneva, Switzerland\\}

\begin{document}
\maketitle

\section{Introduction}
    \begin{wrapfigure}{r}{0.5\textwidth}
    % \begin{figure}[ht]
      \centering
      \includegraphics[width=0.5\textwidth,clip]{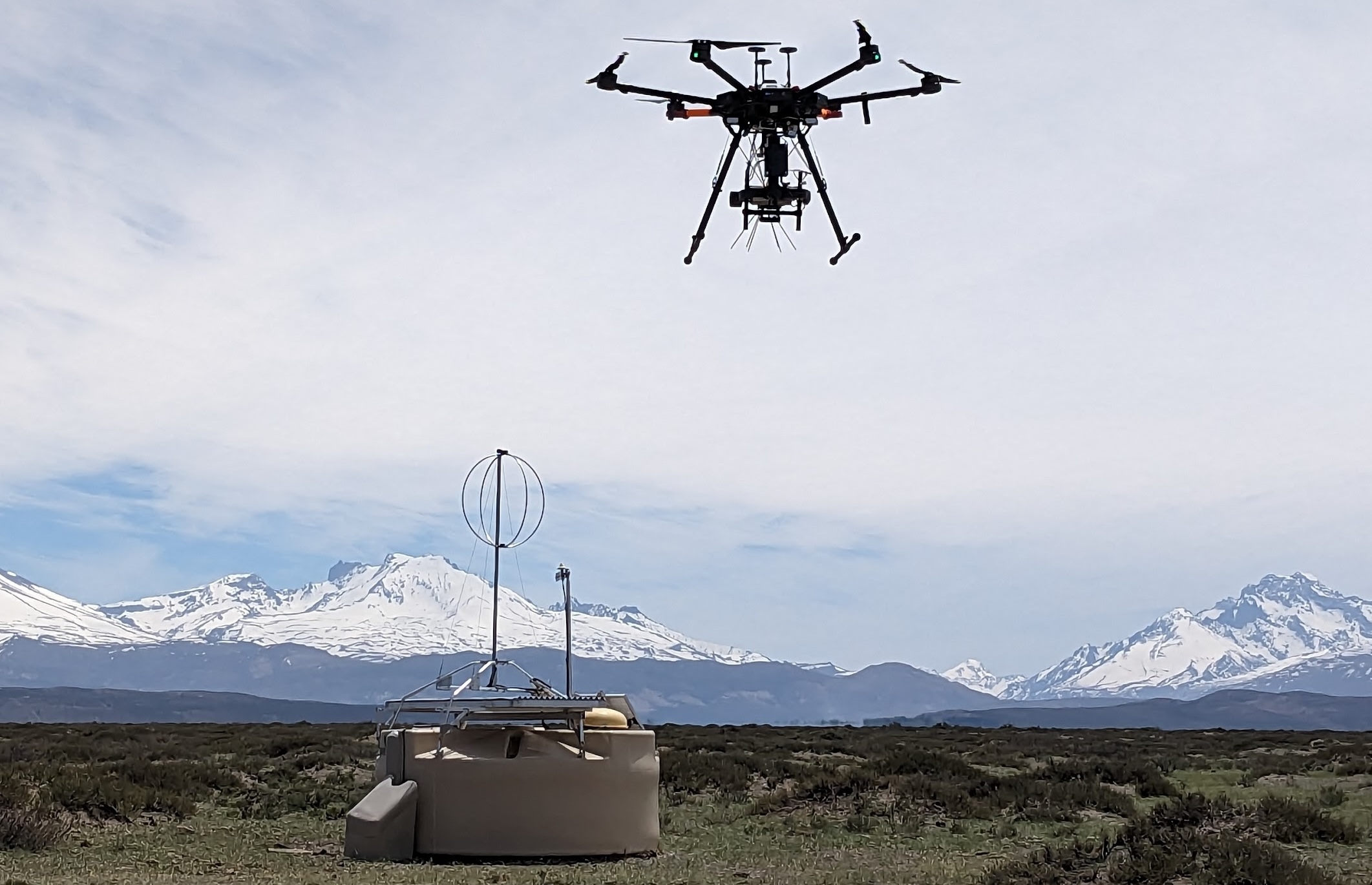}
      \caption{An upgraded surface detector station. The loop antennas on top are part of the Radio Detector. The image was taking during a calibration measurement with a reference radio source mounted on a drone.}
      \label{fig:imageOfRD}
    % \end{figure}
    \end{wrapfigure}
   The Pierre Auger Observatory~\cite{ref:augernim} has recently been enhanced through the AugerPrime upgrade~\cite{ref:AugerPrime}, which aims to improve its capability for studying ultra-high-energy cosmic rays. As part of this upgrade, each water-Cherenkov detector (WCD) in the surface detector array has been complemented with a surface scintillator detector (SSD)~\cite{ref:SSD}, upgraded electronics, and an expanded dynamic range. The Auger Radio Detector (RD) adds two dual-polarized radio antennas at each station to detect the radio emissions from extensive air showers in the $30-80$\,MHz frequency range (see Fig.~\ref{fig:imageOfRD}). The antennas are a revised version of the Short Aperiodic Loaded Loop Antenna (SALLA)~\cite{ref:SALLA}, originally developed for the Auger Engineering Radio Array (AERA)~\cite{ref:AERA_Main}. Radio signals from two antenna channels (aligned along the north-south, and east-west directions) are amplified, filtered, and digitized at $250$\,MHz sampling with 12-bit resolution. Data are collected upon triggers from the WCDs and transferred through a wireless communication system. Future developments may incorporate radio data directly into the trigger decision, potentially enhancing sensitivity to photon and neutrino-induced air showers.
   Given the $1.5$\,km spacing of the surface detector array, this radio extension will enable the detection of inclined air showers, with zenith angles greater than approximately $65^\circ$, over large ground areas. Such showers, as predicted by simulations~\cite{ref:HuegeReview} and observed with AERA~\cite{ref:AERAinclined} and now this detector upgrade, produce detectable radio signals spanning tens of kilometers. For these inclined showers, the electromagnetic component of the shower is largely absorbed before reaching the ground, allowing the WCDs to measure predominantly muonic signals. Conversely, the radio antennas are sensitive only to the electromagnetic cascade.
   This complementary detection approach provides a method to infer the mass composition of primary cosmic rays and serves as an independent cross-check to the WCD–SSD system used for more vertical air showers.
   We present an overview of the Auger Radio Detector deployment, data monitoring and processing, and show the first data.

    \begin{figure}[t]
      \centering
      \includegraphics[width=0.87\textwidth,clip]{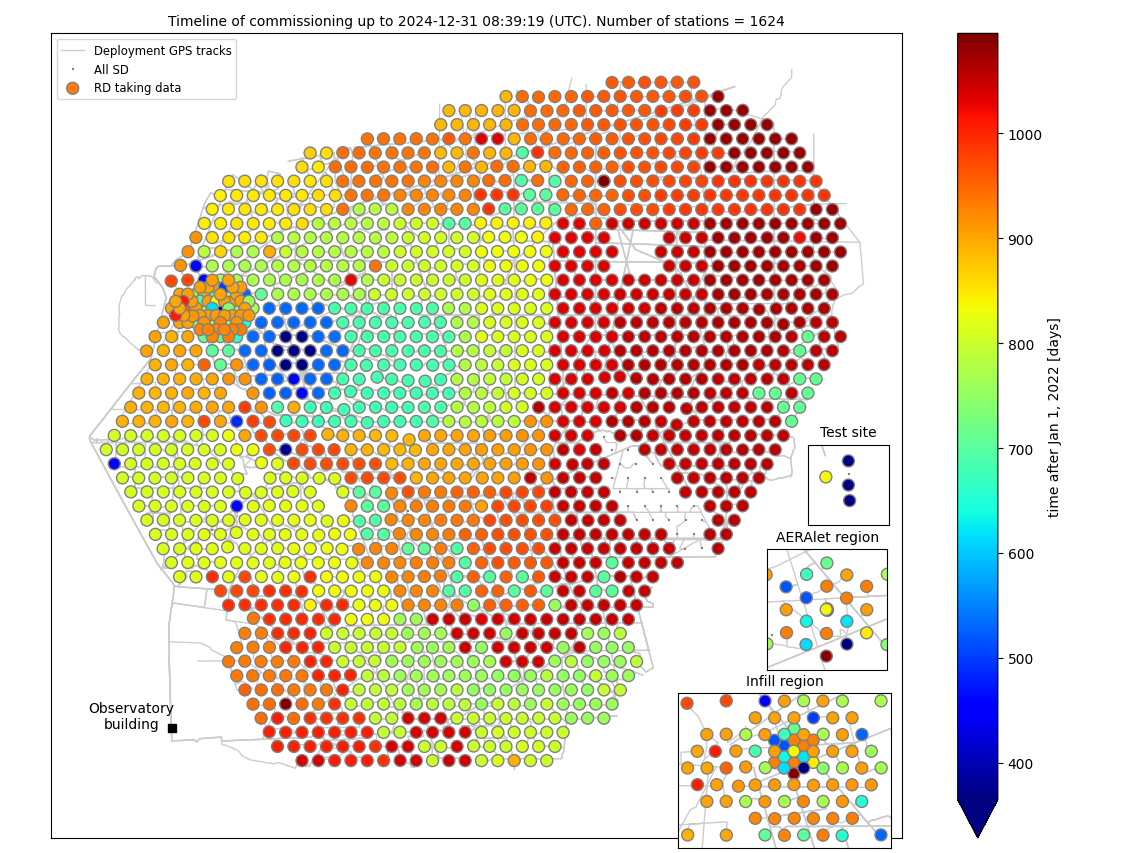}
      \caption{Timeline of the deployment of radio stations from the start of 2022 till the end of 2024. The colors indicate the order of deployment, determined by environmental and logistical constraints. The gray lines mark the GPS tracks of the deployment. The bottom right shows three successive zoom-in panels of the denser region in the north-west region (from bottom to top: station spacings of $750$\,m, $433$\,m, and O($10$)\,m).}
      \label{fig:CommissioningMap}
    \end{figure}
    \begin{figure}[!t]
      \centering
      \includegraphics[width=0.9\textwidth,clip]{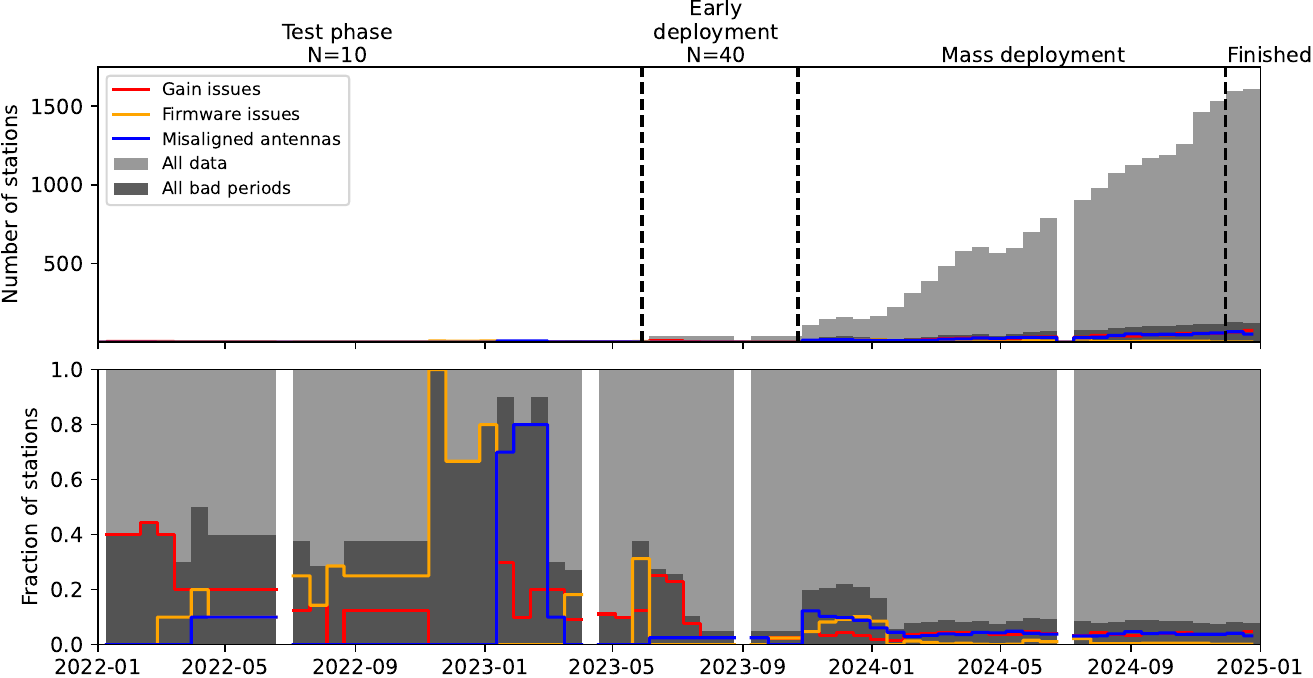}
      \caption{Timeline of the deployment. Several phases of deployment are annotated. Shown are the number of stations in the data acquisition and the number of stations that were flagged during periods of bad operation by an automated data-quality monitoring pipeline. The contributions that are shown: issues with reduced antenna gain (bad connections, broken amplifier, etc.), issues with firmware, and misalignment of antennas.} 
      \label{fig:BadPeriods}
    \end{figure}

\section{Commissioning of the Radio Detector}

    %     \begin{figure}[t]
    % \centering
    % \includegraphics[width=0.49\columnwidth]{img/evt20240420.png}\hspace*{\fill}
    % \includegraphics[width=0.49\columnwidth]{img/evt20241231.png}
    % \caption{RFI map ch0 from MoRD (Left), ch1 (right)}
    % \label{fig:RMSmap_MoRD}
    % \end{figure}
    
    \subsection{Deployment}

    The Radio Detector marks the first extreme-scale deployment of radio antennas in a cosmic-ray detector. Initial testing efforts began in 2019. After this, the deployment proceeded in phases, starting with 10 stations in 2022, then scaling up to 40 in 2023, and starting in November 2023 with the roll-out of the full array. In Fig.~\ref{fig:CommissioningMap} and Fig.~\ref{fig:BadPeriods}, we show the deployment of all radio antennas over time. The roll-out of RD stations was influenced by environmental factors such as the flooding season, extreme heat in sandy regions, snowfall, as well as international shipping delays, logistical constraints, and landowner agreements.  Throughout the process, continuous data monitoring led to design and procedure improvements, including solutions to prevent rotating antennas in strong winds, and static build-up during transport, which had previously damaged some amplifier components.

    \subsection{Data monitoring}
    
    \begin{figure}[t]
      \centering
      \includegraphics[width=0.9\textwidth,clip]{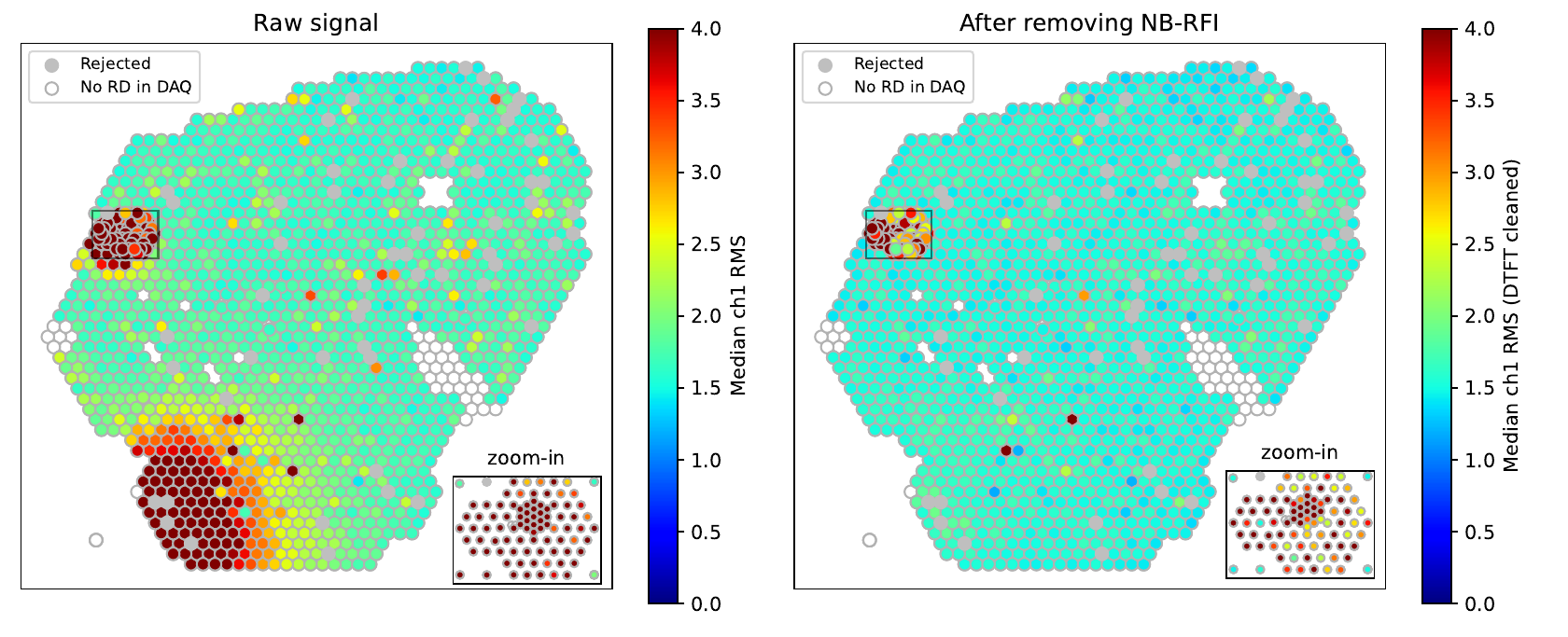}
      \caption{Example of noise the conditions measured by the RD. (Left) Median RMS of the recorded voltage traces of channel 1 (north-south aligned antennas) of each radio station in April 2025. Stations without radio antenna or with bad data quality are marked in white and gray, respectively. (Right) Median RMS of channel 1, like the left figure, but after removing up to $10$ narrow-band emitter signals (NB-RFI) with a discrete time Fourier transform (DTFT). In the south-west, near the single white marker in the bottom left, is the location of the TV channel 4 broadcasting tower that emits an AM signal at $67.25$\,MHz, increasing the RMS in our stations. The increased noise in the zoom-in region is due to additional electronics at these stations. RFI mitigation here has only been completed for selected stations.}
      \label{fig:RMSmap_MoRD}
    \end{figure}
    
    Since early 2023, a live monitoring system has tracked the status and data quality of the Radio Detector. An internal Auger webpage displays key metrics including data availability, software update status, RMS of signal traces, mean frequency spectra, and the most recent voltage traces and frequency spectra. As an example, we show in Fig.~\ref{fig:RMSmap_MoRD} the median RMS noise (root-mean-square of the voltage trace) over the month of April 2025, as measured per station. The live monitoring enables prompt identification of issues such as hardware failures or radio-frequency interference (RFI). Additional checks include detection of repeated traces using hash comparisons, identification of bit flips through unphysical-looking single-bin spikes in the voltage traces, and the analysis of spectral power to detect loss of gain. The latter can indicate failures such as broken low-noise amplifiers, damaged or loose cables, faulty connectors, or dust in connectors. These checks have been implemented in an automated pipeline that identifies periods of bad operation (both for diagnosing issues and filtering out the data before undertaking physics analysis). In Fig.~\ref{fig:BadPeriods} these bad periods are shown. To illustrate the importance of monitoring these quantities in the early deployment phases, one can see that the start of the monitoring website in early 2023 coincides with a sharp drop-off of bad stations due to repairs. Futhermore, at the start of each new deployment phase new issues naturally arise. These can be seen to be resolved promptly by the decrease in the number of bad periods before continuing the large-scale deployment. 
    
    \subsection{Antenna alignment monitoring with reference beacon signals}
    \begin{figure}[t]
      \centering
      \includegraphics[width=0.99\textwidth,clip]{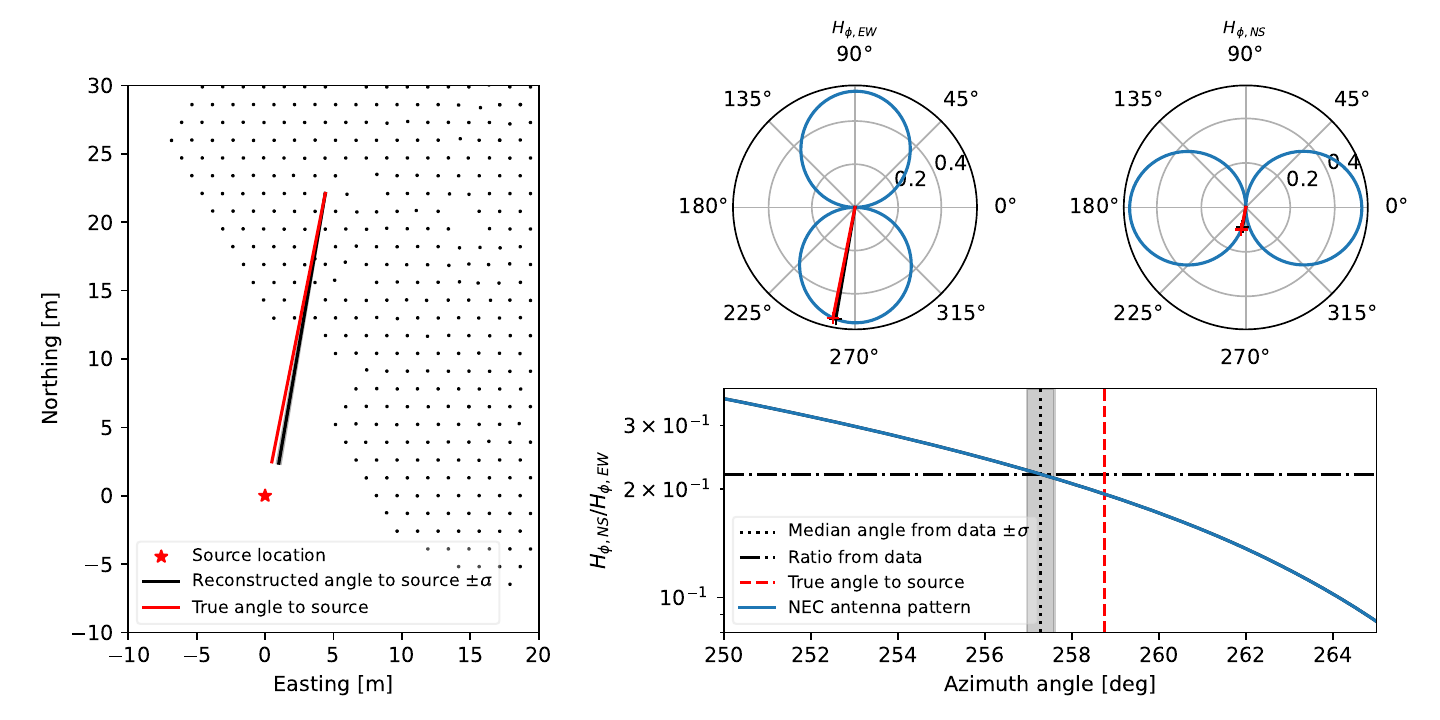}
      \caption{Methodology for measuring antenna alignment. The top two plots on the right-hand side display the simulated antenna sensitivity to vertically polarized signals ($H_{\phi}$) in the East-West and North-South aligned channels. The pattern has been calculated with the NEC software~\cite{ref:UgoARENA2022}. The red line points in the direction of the AM TV signal tower. The black line is the reconstructed direction. In the bottom right plot we compare the ratio of the two channels as predicted by NEC (blue line) and as measured in the data (horizontal line). The intersection provides the estimation of the alignment. Uncertainties are obtained from subtracting the background signal in a bootstrapping procedure. In the left-hand plot, we show the position of the evaluated RD station and the position of the TV signal source.} %ref:4nec2,
      \label{fig:AntennaRotation}
    \end{figure}
    A method was developed to verify station alignment by comparing the measured signal from a nearby TV transmitter (which emits a strong narrow-band signal at $67.25$\,MHz) with the expected response from our antenna model. The procedure is demonstrated and illustrated in Fig.~\ref{fig:AntennaRotation}. This reference signal is visible across the entire array and enables detection of swapped cables (swapping of the two channels) or misaligned stations caused by installation errors or subsequent rotation. An example can be seen in Fig.~\ref{fig:BadPeriods} in November 2023 when extreme wind conditions rotated antennas due to a manufacturing tolerance issue; this was corrected by reinforcing the mounting of the SALLA antennas with an additional rivet. Evaluation of the antenna orientations for the full array shows a small azimuthal dependence of the alignment of a few degrees, hinting at possible deviations in the simulated antenna pattern pointing near the horizon or effects from ground reflection. After removing the azimuthal dependence with a spline fit we obtain a spread in the alignments of less than $5^\circ$ over all stations (which includes both the true misalignment and method uncertainties).

    \subsection{Calibration}
    As an additional high-level check, we validated the absolute gain of our signal chain using the galactic background emission as a reference source~\cite{ref:TomasICRC2021, ref:DiegoARENA2024}. Results show compatible calibration factors to those obtained in an early phase of the deployment~\cite{ref:TomasICRC2021}, demonstrating that the measured signals all over the array are consistent with laboratory-based measurements of the signal chain. In addition, the directional sensitivity was evaluated in 2023 using a drone-mounted calibration source (see Fig.~\ref{fig:imageOfRD}), with preliminary results showing agreement (within systematic uncertainties) with simulations of the antenna response~\cite{ref:AlexARENA2024}. 
    
\section{First Data}
    \begin{figure}[t]
    \centering
    \includegraphics[width=0.99\columnwidth]{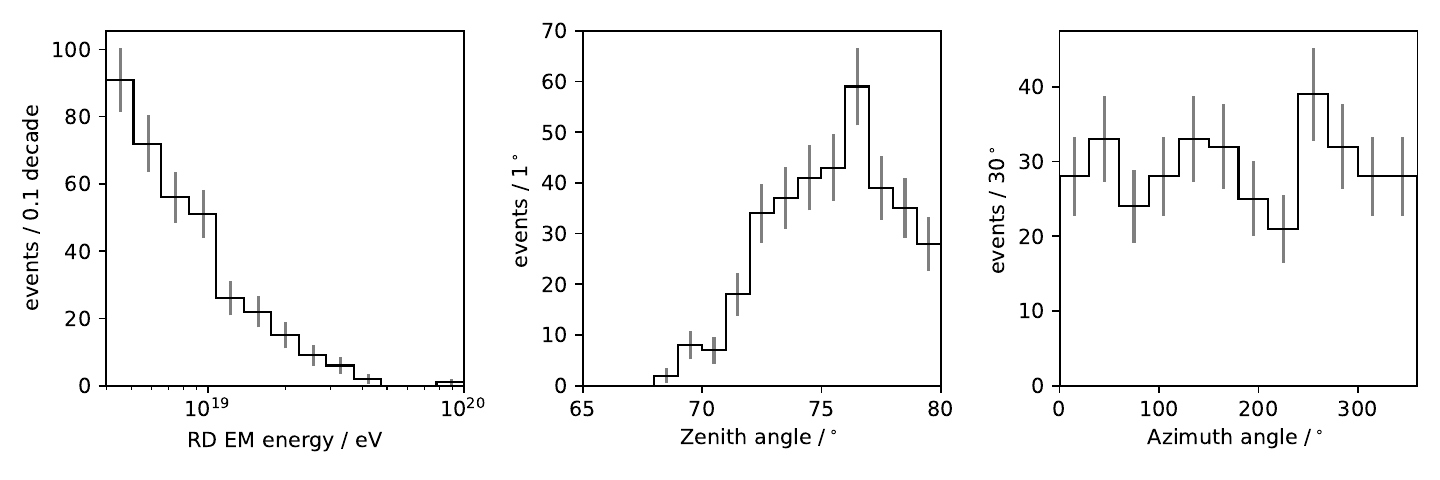}
    \caption{Distribution of reconstructed air-shower parameters, showing the electromagnetic energy (left), the zenith angle (middle), and azimuth angle (right). Poisson uncertainties are shown to illustrate the expected statistical fluctuations. 
    }
    \label{fig:EventHistograms}
    \end{figure}
    
    \subsection{Collected events}
\begin{figure}[t]
    \centering
    \includegraphics[width=0.6\columnwidth]{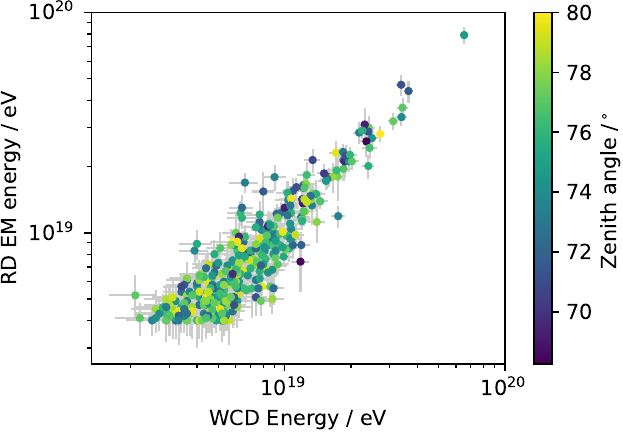}\hspace*{\fill}
    \includegraphics[width=0.4\columnwidth]{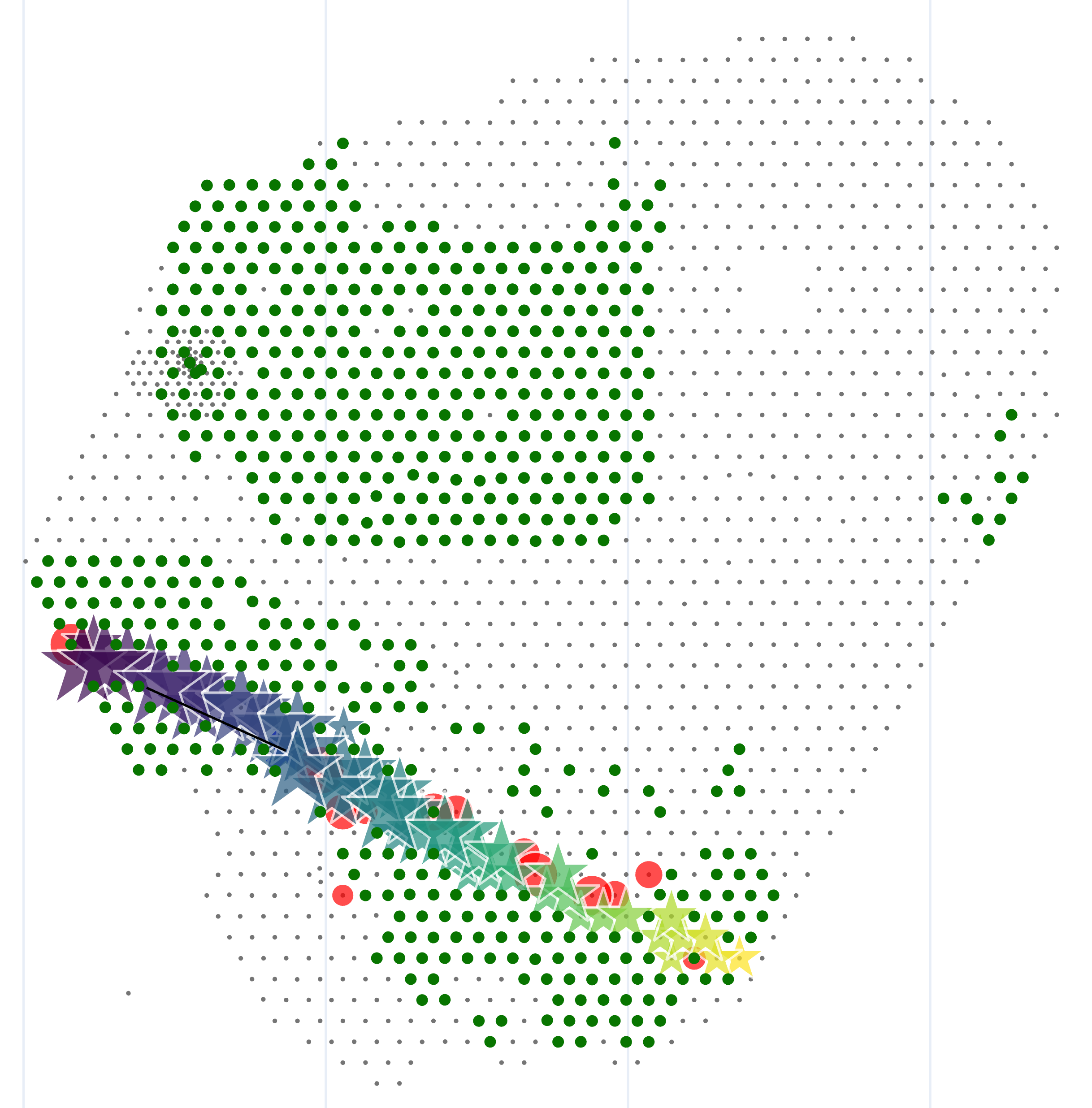}
    \caption{(Left): Comparison of the cosmic ray energy as measured by the WCD and the electromagnetic energy as measured by the RD. Highlighted in color are the zenith angles of the air showers. Statistical uncertainties of reconstructed parameters are shown as gray bars. (Right): An example of a near-horizontal air shower with an energy of approximately $32$\,EeV arriving from west. The gray markers are SD positions and the green markers show where the RD was deployed at the time of the event. Stations that measured a significant radio signal are shown with star markers. The color indicates the arrival time of the pulse. The underlying red markers show the signals measured by the WCD, where the size is proportional to the signal.}
    \label{fig:EventMapAndScatter}
    \end{figure}

    We have analysed data from Jan.~1, 2022 till Mar.~3, 2025 to obtain our 'first data' set of air shower events. Note that the full RD was only deployed for the final three months of this period and the total exposure in this data set is equivalent to approximately half a year of full operation. We applied the bad-period cuts as described in the previous section, rejecting about $6\%$ of raw data.
    We reconstruct air-shower events by identifying radio signals in coincidence with surface detector triggers, calculating the energy fluence in each station, and estimating the incoming direction, shower core position, and electromagnetic energy of the shower using a model of the lateral distribution function; further details are given in~\cite{ref:SimonICRC2025}. We obtain $351$ events above $10^{18.6}$\,eV ($4$\,EeV). In Fig.~\ref{fig:EventHistograms}, we show the distributions of the reconstructed electromagnetic energy, the azimuth angle, and the zenith angle, consisting of $351$ showers after applying all cuts that guarantee good reconstruction of the WCD and RD parameters. Note that the flat distribution in azimuth is as expected due to full efficiency above $4$\,EeV. When going to lower energies we observe the so-called north-south asymmetry caused by the radio signal scaling with the angle to the magnetic field (reducing the number of detectable showers in the direction of the magnetic field axis). 
    In Fig.~\ref{fig:EventMapAndScatter} (Left), we compare the WCD-reconstructed cosmic-ray energy with the radio-reconstructed electromagnetic energy. A good linear correlation can be seen, but note that not all corrections (temperature, absolute calibration, etc.) have yet been implemented and that an inherent offset between the two energies is expected due to the non-electromagnetic energy components, dominated by the invisible energy~\cite{ref:FDInvEnergy}. An independent energy scale from the radio signal by the RD is planned for the future; this will likely follow the detailed procedure established recently for AERA~\cite{ref:TimICRC2025} and extend its result to the highest cosmic-ray energies.

    \subsection{Extremely extended radio footprints}
    In Fig.~\ref{fig:EventMapAndScatter} (Right), we illustrate the detection of a near-horizontal radio footprint measured in early 2024. The air shower had a zenith angle of $85^\circ$ and an electromagnetic energy of about $32$\,EeV, compatible with the estimate of the WCD. 
    This event demonstrates the ability of the RD to measure in this near-horizontal regime, illustrating the potential to measure ultra-high-energy neutral particles. Of particular interest are showers coming from the west where the Andes mountain range provides the potential for earth-skimming neutrinos to interact and to produce an air shower (the event shown does not have a sufficient zenith angle to be considered a candidate).
    
\section{Outlook}
The Radio Detector is expected to operate for at least a decade, providing a substantial increase in the number of cosmic rays with an estimation of the mass~\cite{ref:FelixICRC2021}. Together with the mass measurements by the WCD-SSD at low zenith angles, it will be able to cover most of the southern sky. This will allow for studies of the mass composition of ultra-high-energy cosmic rays. In addition, the measurements of the amount of muons in the shower as a function of energy by this WCD-RD hybrid approach will contribute to addressing the muon puzzle~\cite{ref:MarvinICRC2025}.
In parallel, the Radio Detector will extend the radio-based energy scale to energies beyond $10^{18.5}$\,eV, building on the results and method developed for the AERA radio detector at the Pierre Auger Observatory~\cite{ref:TimICRC2025}. This extension will provide an independent way to access the cosmic-ray energy up to the highest observed energies. 
Furthermore, the use of both amplitude and phase information in the radio signal enables interferometric reconstruction techniques, allowing for a three-dimensional and time-resolved mapping of the air shower emission region~\cite{ref:HarmICRC2025,ref:FelixICRC2021_interferometry}. This capability has been demonstrated to be feasible at AERA, and work is ongoing to implement this for the RD. The interferometric technique offers a new window into the dynamics of the electromagnetic cascade and air shower physics.

\section{Conclusions}
In this contribution, we report on the status and first results from the Radio Detector (RD) at the Pierre Auger Observatory. The RD measures the radio emission from extensive air showers with zenith angles above $65^\circ$\, enabling the reconstruction of their electromagnetic energy. Combined with the muon measurements from the water-Cherenkov detectors (WCDs), this provides sensitivity to cosmic-ray mass composition at the highest energies. Large-scale deployment of the RD began in November~$2023$ and concluded in November~$2024$. A full end-to-end calibration shows agreement between Galactic and laboratory reference measurements to within $5\%$, and continuous system monitoring ensures hardware integrity, including antenna alignment within $5^\circ$. We present first data demonstrating a strong correlation between the RD-reconstructed electromagnetic energy and the total energy measured by the WCDs, confirming the expected performance of the full detector chain—including triggering, readout, calibration, and reconstruction. As an illustration of the capabilities of the RD, an air shower with an energy of $32$\,EeV and a zenith angle of $85^\circ$, producing a $50$\,km radio footprint, was shown. In the next decade, the RD will enable high-statistics composition studies and interferometric reconstruction of the shower development, providing new possibilities for probing the physics of ultra-high-energy air showers and their sources.

\newpage
% \include{latex_authorlist_authors}
% \include{latex_authorlist_institutions}
%% Auger Author List 
\begin{center}
\par\noindent
\textbf{The Pierre Auger Collaboration}
\end{center}
\begin{wrapfigure}[9]{l}{0.12\linewidth}
\vspace{-2.9ex}
\includegraphics[width=0.98\linewidth]{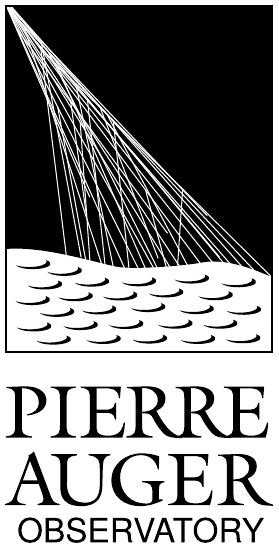}
\end{wrapfigure}
\begin{sloppypar}\noindent
\input{latex_authorlist_authors}
\end{sloppypar}

\vspace{1ex}
\begin{center}
\rule{0.1\columnwidth}{0.5pt}
\raisebox{-0.4ex}{\scriptsize$\bullet$}
\rule{0.1\columnwidth}{0.5pt}
\end{center}

\vspace{1ex}
\input{latex_authorlist_institutions}
\vspace{1ex}
\input{acknowledgments}

%% End Auger Author List

\end{document}

%% file: latex_authorlist_authors.tex
% created on 2025-06-06
A.~Abdul Halim$^{13}$,
P.~Abreu$^{70}$,
M.~Aglietta$^{53,51}$,
I.~Allekotte$^{1}$,
K.~Almeida Cheminant$^{78,77}$,
A.~Almela$^{7,12}$,
R.~Aloisio$^{44,45}$,
J.~Alvarez-Mu\~niz$^{76}$,
A.~Ambrosone$^{44}$,
J.~Ammerman Yebra$^{76}$
% ...

%% file: latex_authorlist_institutions.tex
% created on 2025-06-06
% needs \usepackage{enumitem}
\begin{description}[labelsep=0.2em,align=right,labelwidth=0.7em,labelindent=0em,leftmargin=2em,noitemsep,before={\renewcommand\makelabel[1]{##1 }}]
\item[$^{1}$] Centro At\'omico Bariloche and Instituto Balseiro (CNEA-UNCuyo-CONICET), San Carlos de Bariloche, Argentina
\item[$^{2}$] Departamento de F\'\i{}sica and Departamento de Ciencias de la Atm\'osfera y los Oc\'eanos, FCEyN, Universidad de Buenos Aires and CONICET, Buenos Aires, Argentina
\item[$^{3}$] IFLP, Universidad Nacional de La Plata and CONICET, La Plata, Argentina
\item[$^{4}$] Instituto de Astronom\'\i{}a y F\'\i{}sica del Espacio (IAFE, CONICET-UBA), Buenos Aires, Argentina
\item[$^{5}$] Instituto de F\'\i{}sica de Rosario (IFIR) -- CONICET/U.N.R.\ and Facultad de Ciencias Bioqu\'\i{}micas y Farmac\'euticas U.N.R., Rosario, Argentina
\item[$^{6}$] Instituto de Tecnolog\'\i{}as en Detecci\'on y Astropart\'\i{}culas (CNEA, CONICET, UNSAM), and Universidad Tecnol\'ogica Nacional -- Facultad Regional Mendoza (CONICET/CNEA), Mendoza, Argentina
\item[$^{7}$] Instituto de Tecnolog\'\i{}as en Detecci\'on y Astropart\'\i{}culas (CNEA, CONICET, UNSAM), Buenos Aires, Argentina
\item[$^{8}$] International Center of Advanced Studies and Instituto de Ciencias F\'\i{}sicas, ECyT-UNSAM and CONICET, Campus Miguelete -- San Mart\'\i{}n, Buenos Aires, Argentina
\item[$^{9}$] Laboratorio Atm\'osfera -- Departamento de Investigaciones en L\'aseres y sus Aplicaciones -- UNIDEF (CITEDEF-CONICET), Argentina
\item[$^{10}$] Observatorio Pierre Auger, Malarg\"ue, Argentina
\item[$^{11}$] Observatorio Pierre Auger and Comisi\'on Nacional de Energ\'\i{}a At\'omica, Malarg\"ue, Argentina
\item[$^{12}$] Universidad Tecnol\'ogica Nacional -- Facultad Regional Buenos Aires, Buenos Aires, Argentina
\item[$^{13}$] University of Adelaide, Adelaide, S.A., Australia
\item[$^{14}$] Universit\'e Libre de Bruxelles (ULB), Brussels, Belgium
\item[$^{15}$] Vrije Universiteit Brussels, Brussels, Belgium
\item[$^{16}$] Centro Brasileiro de Pesquisas Fisicas, Rio de Janeiro, RJ, Brazil
\item[$^{17}$] Centro Federal de Educa\c{c}\~ao Tecnol\'ogica Celso Suckow da Fonseca, Petropolis, Brazil
\item[$^{18}$] Instituto Federal de Educa\c{c}\~ao, Ci\^encia e Tecnologia do Rio de Janeiro (IFRJ), Brazil
\item[$^{19}$] Universidade de S\~ao Paulo, Escola de Engenharia de Lorena, Lorena, SP, Brazil
\item[$^{20}$] Universidade de S\~ao Paulo, Instituto de F\'\i{}sica de S\~ao Carlos, S\~ao Carlos, SP, Brazil
\item[$^{21}$] Universidade de S\~ao Paulo, Instituto de F\'\i{}sica, S\~ao Paulo, SP, Brazil
\item[$^{22}$] Universidade Estadual de Campinas (UNICAMP), IFGW, Campinas, SP, Brazil
\item[$^{23}$] Universidade Estadual de Feira de Santana, Feira de Santana, Brazil
\item[$^{24}$] Universidade Federal de Campina Grande, Centro de Ciencias e Tecnologia, Campina Grande, Brazil
\item[$^{25}$] Universidade Federal do ABC, Santo Andr\'e, SP, Brazil
\item[$^{26}$] Universidade Federal do Paran\'a, Setor Palotina, Palotina, Brazil
\item[$^{27}$] Universidade Federal do Rio de Janeiro, Instituto de F\'\i{}sica, Rio de Janeiro, RJ, Brazil
\item[$^{28}$] Universidad de Medell\'\i{}n, Medell\'\i{}n, Colombia
\item[$^{29}$] Universidad Industrial de Santander, Bucaramanga, Colombia
\item[$^{30}$] Charles University, Faculty of Mathematics and Physics, Institute of Particle and Nuclear Physics, Prague, Czech Republic
\item[$^{31}$] Institute of Physics of the Czech Academy of Sciences, Prague, Czech Republic
\item[$^{32}$] Palacky University, Olomouc, Czech Republic
\item[$^{33}$] CNRS/IN2P3, IJCLab, Universit\'e Paris-Saclay, Orsay, France
\item[$^{34}$] Laboratoire de Physique Nucl\'eaire et de Hautes Energies (LPNHE), Sorbonne Universit\'e, Universit\'e de Paris, CNRS-IN2P3, Paris, France
\item[$^{35}$] Univ.\ Grenoble Alpes, CNRS, Grenoble Institute of Engineering Univ.\ Grenoble Alpes, LPSC-IN2P3, 38000 Grenoble, France
\item[$^{36}$] Universit\'e Paris-Saclay, CNRS/IN2P3, IJCLab, Orsay, France
\item[$^{37}$] Bergische Universit\"at Wuppertal, Department of Physics, Wuppertal, Germany
\item[$^{38}$] Karlsruhe Institute of Technology (KIT), Institute for Experimental Particle Physics, Karlsruhe, Germany
\item[$^{39}$] Karlsruhe Institute of Technology (KIT), Institut f\"ur Prozessdatenverarbeitung und Elektronik, Karlsruhe, Germany
\item[$^{40}$] Karlsruhe Institute of Technology (KIT), Institute for Astroparticle Physics, Karlsruhe, Germany
\item[$^{41}$] RWTH Aachen University, III.\ Physikalisches Institut A, Aachen, Germany
\item[$^{42}$] Universit\"at Hamburg, II.\ Institut f\"ur Theoretische Physik, Hamburg, Germany
\item[$^{43}$] Universit\"at Siegen, Department Physik -- Experimentelle Teilchenphysik, Siegen, Germany
\item[$^{44}$] Gran Sasso Science Institute, L'Aquila, Italy
\item[$^{45}$] INFN Laboratori Nazionali del Gran Sasso, Assergi (L'Aquila), Italy
\item[$^{46}$] INFN, Sezione di Catania, Catania, Italy
\item[$^{47}$] INFN, Sezione di Lecce, Lecce, Italy
\item[$^{48}$] INFN, Sezione di Milano, Milano, Italy
\item[$^{49}$] INFN, Sezione di Napoli, Napoli, Italy
\item[$^{50}$] INFN, Sezione di Roma ``Tor Vergata'', Roma, Italy
\item[$^{51}$] INFN, Sezione di Torino, Torino, Italy
\item[$^{52}$] Istituto di Astrofisica Spaziale e Fisica Cosmica di Palermo (INAF), Palermo, Italy
\item[$^{53}$] Osservatorio Astrofisico di Torino (INAF), Torino, Italy
\item[$^{54}$] Politecnico di Milano, Dipartimento di Scienze e Tecnologie Aerospaziali , Milano, Italy
\item[$^{55}$] Universit\`a del Salento, Dipartimento di Matematica e Fisica ``E.\ De Giorgi'', Lecce, Italy
\item[$^{56}$] Universit\`a dell'Aquila, Dipartimento di Scienze Fisiche e Chimiche, L'Aquila, Italy
\item[$^{57}$] Universit\`a di Catania, Dipartimento di Fisica e Astronomia ``Ettore Majorana``, Catania, Italy
\item[$^{58}$] Universit\`a di Milano, Dipartimento di Fisica, Milano, Italy
\item[$^{59}$] Universit\`a di Napoli ``Federico II'', Dipartimento di Fisica ``Ettore Pancini'', Napoli, Italy
\item[$^{60}$] Universit\`a di Palermo, Dipartimento di Fisica e Chimica ''E.\ Segr\`e'', Palermo, Italy
\item[$^{61}$] Universit\`a di Roma ``Tor Vergata'', Dipartimento di Fisica, Roma, Italy
\item[$^{62}$] Universit\`a Torino, Dipartimento di Fisica, Torino, Italy
\item[$^{63}$] Benem\'erita Universidad Aut\'onoma de Puebla, Puebla, M\'exico
\item[$^{64}$] Unidad Profesional Interdisciplinaria en Ingenier\'\i{}a y Tecnolog\'\i{}as Avanzadas del Instituto Polit\'ecnico Nacional (UPIITA-IPN), M\'exico, D.F., M\'exico
\item[$^{65}$] Universidad Aut\'onoma de Chiapas, Tuxtla Guti\'errez, Chiapas, M\'exico
\item[$^{66}$] Universidad Michoacana de San Nicol\'as de Hidalgo, Morelia, Michoac\'an, M\'exico
\item[$^{67}$] Universidad Nacional Aut\'onoma de M\'exico, M\'exico, D.F., M\'exico
\item[$^{68}$] Institute of Nuclear Physics PAN, Krakow, Poland
\item[$^{69}$] University of \L{}\'od\'z, Faculty of High-Energy Astrophysics,\L{}\'od\'z, Poland
\item[$^{70}$] Laborat\'orio de Instrumenta\c{c}\~ao e F\'\i{}sica Experimental de Part\'\i{}culas -- LIP and Instituto Superior T\'ecnico -- IST, Universidade de Lisboa -- UL, Lisboa, Portugal
\item[$^{71}$] ``Horia Hulubei'' National Institute for Physics and Nuclear Engineering, Bucharest-Magurele, Romania
\item[$^{72}$] Institute of Space Science, Bucharest-Magurele, Romania
\item[$^{73}$] Center for Astrophysics and Cosmology (CAC), University of Nova Gorica, Nova Gorica, Slovenia
\item[$^{74}$] Experimental Particle Physics Department, J.\ Stefan Institute, Ljubljana, Slovenia
\item[$^{75}$] Universidad de Granada and C.A.F.P.E., Granada, Spain
\item[$^{76}$] Instituto Galego de F\'\i{}sica de Altas Enerx\'\i{}as (IGFAE), Universidade de Santiago de Compostela, Santiago de Compostela, Spain
\item[$^{77}$] IMAPP, Radboud University Nijmegen, Nijmegen, The Netherlands
\item[$^{78}$] Nationaal Instituut voor Kernfysica en Hoge Energie Fysica (NIKHEF), Science Park, Amsterdam, The Netherlands
\item[$^{79}$] Stichting Astronomisch Onderzoek in Nederland (ASTRON), Dwingeloo, The Netherlands
\item[$^{80}$] Universiteit van Amsterdam, Faculty of Science, Amsterdam, The Netherlands
\item[$^{81}$] Case Western Reserve University, Cleveland, OH, USA
\item[$^{82}$] Colorado School of Mines, Golden, CO, USA
\item[$^{83}$] Department of Physics and Astronomy, Lehman College, City University of New York, Bronx, NY, USA
\item[$^{84}$] Michigan Technological University, Houghton, MI, USA
\item[$^{85}$] New York University, New York, NY, USA
\item[$^{86}$] University of Chicago, Enrico Fermi Institute, Chicago, IL, USA
\item[$^{87}$] University of Delaware, Department of Physics and Astronomy, Bartol Research Institute, Newark, DE, USA
\item[] -----
\item[$^{a}$] Max-Planck-Institut f\"ur Radioastronomie, Bonn, Germany
\item[$^{b}$] also at Kapteyn Institute, University of Groningen, Groningen, The Netherlands
\item[$^{c}$] School of Physics and Astronomy, University of Leeds, Leeds, United Kingdom
\item[$^{d}$] Fermi National Accelerator Laboratory, Fermilab, Batavia, IL, USA
\item[$^{e}$] Pennsylvania State University, University Park, PA, USA
\item[$^{f}$] Colorado State University, Fort Collins, CO, USA
\item[$^{g}$] Louisiana State University, Baton Rouge, LA, USA
\item[$^{h}$] now at Graduate School of Science, Osaka Metropolitan University, Osaka, Japan
\item[$^{i}$] Institut universitaire de France (IUF), France
\item[$^{j}$] now at Technische Universit\"at Dortmund and Ruhr-Universit\"at Bochum, Dortmund and Bochum, Germany
\end{description}

%% file: acknowledgments.tex
% created on 2025-06-06
\section*{Acknowledgments}

\begin{sloppypar}
The successful installation, commissioning, and operation of the Pierre
Auger Observatory would not have been possible without the strong
commitment and effort from the technical and administrative staff in
Malarg\"ue. We are very grateful to the following agencies and
organizations for financial support:
\end{sloppypar}

\begin{sloppypar}
Argentina -- Comisi\'on Nacional de Energ\'\i{}a At\'omica; Agencia Nacional de
Promoci\'on Cient\'\i{}fica y Tecnol\'ogica (ANPCyT); Consejo Nacional de
Investigaciones Cient\'\i{}ficas y T\'ecnicas (CONICET); Gobierno de la
Provincia de Mendoza; Municipalidad de Malarg\"ue; NDM Holdings and Valle
Las Le\~nas; in gratitude for their continuing cooperation over land
access; Australia -- the Australian Research Council; Belgium -- Fonds
de la Recherche Scientifique (FNRS); Research Foundation Flanders (FWO),
Marie Curie Action of the European Union Grant No.~101107047; Brazil --
Conselho Nacional de Desenvolvimento Cient\'\i{}fico e Tecnol\'ogico (CNPq);
Financiadora de Estudos e Projetos (FINEP); Funda\c{c}\~ao de Amparo \`a
Pesquisa do Estado de Rio de Janeiro (FAPERJ); S\~ao Paulo Research
Foundation (FAPESP) Grants No.~2019/10151-2, No.~2010/07359-6 and
No.~1999/05404-3; Minist\'erio da Ci\^encia, Tecnologia, Inova\c{c}\~oes e
Comunica\c{c}\~oes (MCTIC); Czech Republic -- GACR 24-13049S, CAS LQ100102401,
MEYS LM2023032, CZ.02.1.01/0.0/0.0/16{\textunderscore}013/0001402,
CZ.02.1.01/0.0/0.0/18{\textunderscore}046/0016010 and
CZ.02.1.01/0.0/0.0/17{\textunderscore}049/0008422 and CZ.02.01.01/00/22{\textunderscore}008/0004632;
France -- Centre de Calcul IN2P3/CNRS; Centre National de la Recherche
Scientifique (CNRS); Conseil R\'egional Ile-de-France; D\'epartement
Physique Nucl\'eaire et Corpusculaire (PNC-IN2P3/CNRS); D\'epartement
Sciences de l'Univers (SDU-INSU/CNRS); Institut Lagrange de Paris (ILP)
Grant No.~LABEX ANR-10-LABX-63 within the Investissements d'Avenir
Programme Grant No.~ANR-11-IDEX-0004-02; Germany -- Bundesministerium
f\"ur Bildung und Forschung (BMBF); Deutsche Forschungsgemeinschaft (DFG);
Finanzministerium Baden-W\"urttemberg; Helmholtz Alliance for
Astroparticle Physics (HAP); Helmholtz-Gemeinschaft Deutscher
Forschungszentren (HGF); Ministerium f\"ur Kultur und Wissenschaft des
Landes Nordrhein-Westfalen; Ministerium f\"ur Wissenschaft, Forschung und
Kunst des Landes Baden-W\"urttemberg; Italy -- Istituto Nazionale di
Fisica Nucleare (INFN); Istituto Nazionale di Astrofisica (INAF);
Ministero dell'Universit\`a e della Ricerca (MUR); CETEMPS Center of
Excellence; Ministero degli Affari Esteri (MAE), ICSC Centro Nazionale
di Ricerca in High Performance Computing, Big Data and Quantum
Computing, funded by European Union NextGenerationEU, reference code
CN{\textunderscore}00000013; M\'exico -- Consejo Nacional de Ciencia y Tecnolog\'\i{}a
(CONACYT) No.~167733; Universidad Nacional Aut\'onoma de M\'exico (UNAM);
PAPIIT DGAPA-UNAM; The Netherlands -- Ministry of Education, Culture and
Science; Netherlands Organisation for Scientific Research (NWO); Dutch
national e-infrastructure with the support of SURF Cooperative; Poland
-- Ministry of Education and Science, grants No.~DIR/WK/2018/11 and
2022/WK/12; National Science Centre, grants No.~2016/22/M/ST9/00198,
2016/23/B/ST9/01635, 2020/39/B/ST9/01398, and 2022/45/B/ST9/02163;
Portugal -- Portuguese national funds and FEDER funds within Programa
Operacional Factores de Competitividade through Funda\c{c}\~ao para a Ci\^encia
e a Tecnologia (COMPETE); Romania -- Ministry of Research, Innovation
and Digitization, CNCS-UEFISCDI, contract no.~30N/2023 under Romanian
National Core Program LAPLAS VII, grant no.~PN 23 21 01 02 and project
number PN-III-P1-1.1-TE-2021-0924/TE57/2022, within PNCDI III; Slovenia
-- Slovenian Research Agency, grants P1-0031, P1-0385, I0-0033, N1-0111;
Spain -- Ministerio de Ciencia e Innovaci\'on/Agencia Estatal de
Investigaci\'on (PID2019-105544GB-I00, PID2022-140510NB-I00 and
RYC2019-027017-I), Xunta de Galicia (CIGUS Network of Research Centers,
Consolidaci\'on 2021 GRC GI-2033, ED431C-2021/22 and ED431F-2022/15),
Junta de Andaluc\'\i{}a (SOMM17/6104/UGR and P18-FR-4314), and the European
Union (Marie Sklodowska-Curie 101065027 and ERDF); USA -- Department of
Energy, Contracts No.~DE-AC02-07CH11359, No.~DE-FR02-04ER41300,
No.~DE-FG02-99ER41107 and No.~DE-SC0011689; National Science Foundation,
Grant No.~0450696, and NSF-2013199; The Grainger Foundation; Marie
Curie-IRSES/EPLANET; European Particle Physics Latin American Network;
and UNESCO.
\end{sloppypar}

%% file: main.bbl
{\small
\begingroup

\begin{thebibliography}{10}
\bibitem{ref:augernim}
A. Aab \emph{et al.} (Pierre Auger Collaboration), \href{https://doi.org/10.1016/j.nima.2015.06.058}
     {\emph{Nucl. Instrum. Meth. A} \textbf{798} (2015) 172--213}

\bibitem{ref:AugerPrime}
A. Castellina (Pierre Auger Collaboration), \href{https://doi.org/10.1051/epjconf/201921006002}{\emph{EPJ Web Conf.} \textbf{210} (2019) 06002}

\bibitem{ref:SSD}
G. Cataldi (Pierre Auger Collaboration), \href{https://doi.org/10.22323/1.395.0251}{PoS(ICRC2021)251}

\bibitem{ref:SALLA}
O. Krömer, H. Gemmeke, W.D. Apel, \emph{et al.}  \href{https://publikationen.bibliothek.kit.edu/1000178459}{PoS(ICRC2009)1232}

\bibitem{ref:AERA_Main}
P. Abreu \emph{et al.} (Pierre Auger Collaboration), \href{http://dx.doi.org/10.1088/1748-0221/7/10/P10011}
     {\emph{J. Instrum.} \textbf{7} (2012) P10011}

\bibitem{ref:HuegeReview}
T. Huege, \href{https://doi.org/10.1016/j.physrep.2016.02.001}{Phys. Rept. 620, 1--52 (2016)}

\bibitem{ref:AERAinclined}
A. Aab \emph{et al.} (Pierre Auger Collaboration), \href{https://doi.org/10.1088/1475-7516/2018/10/026}
     {\emph{JCAP} \textbf{10} (2018) 026}

% \bibitem{ref:4nec2}
% A. Voors, \href{https://www.qsl.net/4nec2/}{4NEC2 software}.

\bibitem{ref:UgoARENA2022}
U. Giaccari (Pierre Auger Collaboration), \href{https://doi.org/10.22323/1.424.0042}{PoS(ARENA2022)042}

\bibitem{ref:TomasICRC2021}
T. Fodran (Pierre Auger Collaboration), \href{https://doi.org/10.22323/1.395.0262}{PoS(ICRC2021)262}

\bibitem{ref:DiegoARENA2024}
D. Correia dos Santos (Pierre Auger Collaboration), \href{https://doi.org/10.22323/1.470.0030}{PoS(ARENA2024)030}

\bibitem{ref:AlexARENA2024}
A. Reuzki (Pierre Auger Collaboration), \href{https://doi.org/10.22323/1.470.0029}{PoS(ARENA2024)029}

\bibitem{ref:SimonICRC2025}
S. Str\"ahnz (Pierre Auger Collaboration), PoS(ICRC2025)445

\bibitem{ref:TimICRC2025}
T. Huege (Pierre Auger Collaboration), PoS(ICRC2025)616


\bibitem{ref:FelixICRC2021}
F. Schl\"uter (Pierre Auger Collaboration), \href{https://doi.org/10.22323/1.395.0262}{PoS(ICRC2021)262}

\bibitem{ref:FDInvEnergy}
A. Aab \emph{et al.} (Pierre Auger Collaboration),
\href{https://doi.org/10.1103/PhysRevD.100.082003}{Phys. Rev. D \textbf{100}, 082003 (2019)}

\bibitem{ref:MarvinICRC2025}
M. Gottowik (Pierre Auger Collaboration), PoS(ICRC2025)417

\bibitem{ref:HarmICRC2025}
H. Schoorlemmer (Pierre Auger Collaboration), PoS(ICRC2025)297

\bibitem{ref:FelixICRC2021_interferometry}
F. Schl\"uter (Pierre Auger Collaboration), \href{https://doi.org/10.22323/1.395.0228}{PoS(ICRC2021)228}

\end{thebibliography}
\endgroup
}
